# MESOSCOPIC COLLECTIVE ACTIVITY IN EXCITATORY NEURAL FIELDS: GOVERNING EQUATIONS

ABSTRACT. In this study we derive the governing equations for mesoscopic collective activity in the cortex, starting from the generic Hodgkin-Huxley equations for microscopic cell dynamics. For simplicity, and to maintain focus on the essential elements of the derivation, the discussion is confined to excitatory neural fields. The fundamental assumption of the procedure is that mesoscale processes are macroscopic with respect to cell-scale activity, and emerge as the average behavior of a large population of cells. Because of their duration, action-potential details are assumed not observable at mesoscale; the essential mesoscopic function of action potentials is assumed to redistribute energy in the neural field. The Hodgkin-Huxley dynamical model is first reduced to a set of equations that describe subthreshold dynamics. An ensemble average over a cell population then produces a closed system of equations involving two mesoscopic state variables: the density of kinetic energy $J$, carried by sodium ionic currents, and the excitability $H$ of the neural field, which could be described as the average state of gating variable $h$. The resulting model shares some similarities with the Wilson-Cowan class of models, but because the equations are derived directly from microscopic dynamics, it also exhibits significant differences: mesoscopic activity is represented as essentially a subthreshold process; and the dynamical role of the firing rate is naturally reassessed as describing energy transfers (process variable). The linear properties of the equations are consistent with expectations for the dynamics of excitatory neural fields: the system supports oscillations of progressive waves, with shorter waves typically having higher frequencies, propagating slower, and decaying faster. Some nonlinear properties of the equations, with relevance to cross frequency coupling (e.g. theta-gamma), are analyzed in a companion paper. Extending the derivation to include more complex cell dynamics (e.g., including other ionic channels, e.g., calcium channels) and multiple-type, excitatory-inhibitory, neural fields is straightforward, and will be presented elsewhere.

Short running title: Collective Neural Activity: Governing Equations

Authors: Y. Qin[1] and A. Sheremet[1,2]

Affiliations: 1. Engineering School of Sustainable Infrastructure and Environment, University of Florida, Gainesville, FL. 32611.

2. McKnight Brain Institute, Department of Neuroscience, University of Florida, Gainesville, FL. 32610.

Correspondence: Alex Sheremet, email: alex.sheremet@essie.ufl.edu

Competing Interests: The authors declare that they have no competing interests.

Data Availability: No laboratory data was used in the research presented in this manuscript.

Funding information: This work was supported by the McKnight Brain Research Foundation, National Institute on Aging, grant number AG055544, and National Institute of Mental Health, grant number MH109548.

CONTENTS

# 1. INTRODUCTION

Three scales of neural activity are commonly identified in the cortex: the cell (smallest) scale, where activity involves a small number of neurons; the global (largest) scale which encompasses more than one cortical areas; and a third scale, the mesoscale, intermediate between cell and global scales. Previous studies [e.g., Freeman, 1959, Green and Arduini, 1954, Buzsáki et al., 1992, Freeman, 2000b, Muller et al., 2018, and many others], define the mesoscale by activity patterns spanning mm of cortex, with time scales in the order of 10 ms – corresponding to the gamma rhythm, prominent in the hippocampus [Bragin et al., 1995]. While it is generally agreed that cell scale processes are the foundation of the brain activity, and global scale ones are related to the integration of activity across cortex regions, the significance of the mesoscopic activity is still being debated. It seems to be intimately related to the cortex, where it is ubiquitous[1], and typically takes the form of spatio-temporal patterns, such as propagating waves [Petsche and Stumpf, 1960, Lubenov and Siapas, 2009, Patel et al., 2012, 2013, Muller et al., 2018], that appear to develop as perturbations of a largely scale-free background state with power law LFP frequency spectra, [Sheremet et al., 2016, 2019]). Following Freeman [2000a], we will refer to mesoscopic spatio-temporal patterns of neural activity as "collective activity"[2].

A note on terminology. Below, we use the words "cell" and "global" instead of the more common neuroscience "micro/macro" scale names [Muller et al., 2018]. The reason is to avoid confusion with the physics usage. In physics, processes with characteristic lengths $\lambda_1$ and $\lambda_2$ are considered in a micro/macro relation if $\lambda_1 \ll \lambda_m$, where $\ll$ means "much smaller than"; the micro/macro duality is therefore just an order relation. Characteristic scales are used to separate, for example, Brownian motion processes into microscopic (particle collisions at lengths $\lambda_\mu \approx 10^{-4}$ cm) and macroscopic (diffusion process, at resolutions $\lambda_m \approx 1$ cm). However, unless $\lambda_1$ and $\lambda_2$ are orders of magnitude apart, the exact meaning of "much smaller than" is debatable. The meaningful scale distinction comes from distinct governing laws. In the case of Brownian motion, microscopic dynamics are governed by the equations of particle collision, while macroscopic processes are governed by the diffusion equation. Applied to the problem of collective activity, mesoscale is a distinct scale inasmuch as its physical laws are different from both cell and global scales. This then implies that it is macroscopic with respect to cell scale, and microscopic with respect to the global scale.

The physical support of mesoscopic collective activity is not clear, and models can vary. A well studied class of models is based on the Kuramoto's seminal work (e.g., Kuramoto, 1975, Strogatz, 2000, and many others). In this representTation, the central element of neuron dynamics is the action potential (e.g., theta-neuron model the Ermentrout-Kopell canonical model Ermentrout and Kopell, 1986, Byrne et al., 2019, 2020, and many others): neurons are described as weakly-interacting nonlinear oscillators, continuously, metronome-like cycling through action-potential, refractory, and resting phases. Their interaction is weak in the sense that it can only affect the phase (Ermentrout-Kopell canonical model, Ermentrout, 2008), and collective activity arises as a process of synchronization of neural populations. These ideas have been widely applied to analyzing brain activity, for example, to the study of cross-frequency coupling in the LFP recordings [e.g., Guevara and Glass, 1982, Tass et al., 1998, Belluscio et al., 2012, and many others].

---

[1]...in the olfactory bulb Freeman (1978); visual cortex, Prechtl et al. (1997); rodent barrel cortex, Buzsaki and Dragoi (2006); hippocampus, Lubenov and Siapas (2009), Petsche and Stumpf (1960), Muller et al. (2018); auditory cortex Song et al., (2006); motor cortex, Rubino et al. (2006); parietal cortex, Takagaki et al. (2008); across the entire human cortex during sleep, Massimini et al. (2004).

[2]The "mesoscopic collective activity" concept is identical to Freeman's [1975] "mass action". We prefer "collective action " because the word "mass" has a reserved meaning in physics. Freeman [e.g., 2000a,b] suggested that collective activity might hold the key to cognition (and idea consistent with the concept of degeneracy proposed by Edelman and Gally, 2001).

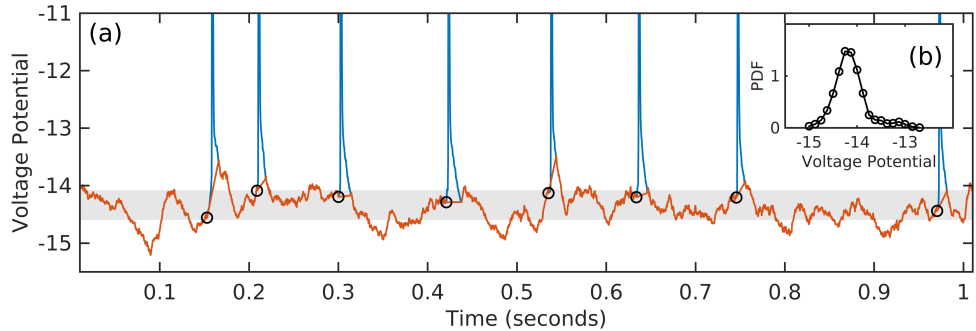

FIGURE 1. a) Example of membrane potential evolution of a pyramidal neuron in CA1; 1-s segment intracellular recording data (blue, courtesy of English et al., 2014; to zoom into the subthreshold regime, the tip of action potentials is cut off); orange line – subthreshold fluctuations with action potentials removed. Circles mark the initiation of an action potential. The gray rectangle marks the vertical span of threshold potential for the 1-s time series shown. The probabilistic width is in fact larger: b) PDF function of the threshold potentials computed using the entire recording. Total recording time was 3059 seconds; total time spent by neuron in action potentials was 326 seconds; the neuron fired 25907 times, corresponding to an average firing rate of 8.49 Hz.

However, at least some observational data suggest an alternative view. In intracellular recordings, rather than continuously cycling (i.e., firing), cells appear to spend most of the time in a "subthreshold regime", with membrane potential fluctuating randomly below the threshold of action potential initiation. Firing events are also infrequent and random. This behavior contradicts the fundamental tenet of the synchronization model, and thus suggests that at least for some parts of the brain, both the questions of the physical support of collective activity, and of its role in cognition remain open.

The observation that cell-scale dynamics are stochastic is not new, and is consistent with the stochastic character of cortex connectivity. Citing from Muller et al. [2018]: "... neurons are embedded in a sea of so-called spontaneous activity. Pyramidal cells in V1 have dense recurrent synaptic connectivity, each receiving from 6,000 to 13,000 inputs per cell. Roughly 80% of these connections come from within the same cortical area, 95% of which arise from neurons within 2 mm of the receiving cell." One may add that synapses have a life-span of a few weeks [Attardo et al., 2015, Holtmaat et al., 2005, Xu et al., 2009, Xiao et al., 2009]; mossy fibers from a granule neuron have up to 200 different synaptic inputs onto a wide variety of neurons [Amaral et al., 2007]. Freeman [2000b] proposed the name "stochastic chaos", "because it arises from and feeds on the randomized activity of myriads of neurons, and it provides the basis for self-organization."

This study makes the fundamental assumption that, as a macroscopic expression of stochastic cell-scale processes, mesoscopic collective activity emerges as averaged behavior of cell populations, and therefore it is insensitive to the precise dynamics of the individual cell. This approach is also not new. The key governing laws of collective activity based on the stochastic chaos assumption were first derived by Wilson and Cowan [1972, 1973] and further refined by Amari, 1975, 1977, Wright and Liley, 1995, Jirsa and Haken, 1996, 1997, Robinson et al., 1997, Cowan et al., 2016 and many others (see, e.g., reviews by Deco et al., 2008, Coombes et al., 2014, Cowan et al., 2016; we will refer collectively to models sharing this approach as the Wilson-Cowan – WC – class of models). While it is hard to overstate how consequential WC models have been for the development of neuroscience and computing science, it is important to note that, at least originally, they were motivated by the search for a reliable computation paradigm based on "unreliable elements" rather than a model for brain activity (e.g., Vinograd and Cowan, 1963; see also the insightful historical account in Anderson and Rosenfeld, 2000). As such they carry a legacy of simplifying assumptions that were rarely explicitly derived and continuously being corrected and improved upon. For example, WC models were originally formulated as a relationship between the local firing rate and excitation pulses received in a given area of the network. Using the (measurable) firing rate as a state variable is a rational choice, but is physically inconsistent, because the firing rate a *process* variable: it describes system energy exchanges, rather the its state (see energy balance discussion in section 2). In general, process variables cannot be used to define the state of the system if there are more than one state variables, and/or if their relation to the state variable is nonlinear[3]. Amari's [1975, 1977] introduction of the averaged membrane potential corrected this issue, but additional variables might be needed. Indeed, action potentials have two major consequences: energy is released into network, but the ability of the network to respond to input is reduced (refractoriness). A measure of the averaged refractoriness of the network seems essential, as it plays an critical role in the formation of mesoscopic patterns [Curtu and Ermentrout, 2001, Meijer and Coombes, 2014], by constraining the local response of the network to incoming excitation.

Most WC-class mesoscopic neural field models were postulated as phenomenological models, i.e., based on some interpretation of observable (measurable) mesoscopic behavior (see, e.g. a comprehensive review by Terry et al., 2022). While phenomenological models provide an effective macroscopic description of multi-scales neural system, Qin et al. [2020] argue that a rigorous derivation should follow standard steps taken in physics to connect microscopic and macroscopic models of multi-scale systems, involving statistical, kinetic, thermodynamic, and hydrodynamic representations (see, e.g., Kardar, 2007b,a). Perhaps the only example of applying such an approach is provided by the series of papers by Byrne et al. [2019, 2020], which derive macroscopic equations for theta-neuron populations starting from the cell-scale Ermentrout-Kopell canonical model. Here, we follow the ideas and assumptions of Qin et al. [2020], and derive mesoscopic equations by scaling up from the standard Hodgkin and Huxley [1952] cell-scale dynamics. For simplicity, we demonstrate the approach for excitatory neural fields.

---

[3]In thermodynamics, temperature $T$ is a state variable, uniquely defined at equilibrium states; the heat $Q$ is a process variable, defined only for state changes. Although related to $T$, $Q$ does not even define uniquely changes $T$. For example, if the water is liquid, receiving $Q$ may cause $T$ to increase by $\Delta T$; but, if the water is ice, the increase would be $\approx 2\Delta T$ (the specific heat capacity of liquid water is twice that of ice); or $\Delta T = 0$, if water is a mixture of liquid and ice, and the incoming energy is used to break molecular bonds and melt some of the ice mass.

The derivation of the collective activity model is presented in section (2). In section (3), we linearize the governing equations and discuss the dependence of its stability on input and characteristic parameters of the neural field. The results are summarized and discussed in section (4).

## 2. DERIVATION OF THE GOVERNING EQUATIONS FOR COLLECTIVE ACTIVITY

A rigorous derivation of the governing equations for collective activity should include a procedure for scaling up physical laws from cell-scale to mesoscale. Microscopic dynamics involve a very large number of degrees of freedom (e.g., in a gas: positions and velocities of molecules), associated with variables that are not observable at macroscale. Scaling-up procedures developed in statistical mechanics [e.g., Kardar, 2007a,b], reduce these to a few macroscopic variables (e.g., in a gas: temperature, pressure, volume, etc) through steps that include the derivation of a kinetic equation for distributions of microscopic variables [e.g., Boltzmann, 1872, Alexeev, 2004, Boltzmann, 2003], ensemble averaging, and hydrodynamic approximations [e.g., Tong, 2012, Kardar, 2007b,a].

Neuronal dynamics is similarly characterized at cell-scale by large number of degrees of freedom (e.g., membrane potential and gating variables of each cell Hodgkin and Huxley, 1952). Mesoscopic collective activity (macroscopic with respect to cell scale) should similarly be described by a few variables.

The scaling up procedure outlined below is heuristic. A rigorous scaling up approach, going through the steps enumerated above, is not available at this time, and would be outside the scope of this study. However, the lack of one should not deter. As Kardar [2007a] notes, "a step-by-step derivation of the macroscopic properties from the microscopic equations of motion is generally impossible, and largely unnecessary". Just like the thermodynamic theory of gases, whose development preceded that of statistical mechanics, macroscopic WC models have been used successfully for several decades now without a rigorous derivation from cell scale dynamics. However, a brief inspection, albeit heuristic, of how collective activity laws might arise from microscopic dynamics might help identify a minimal, complete set of state variables for mesoscopic activity.

Because the cell-scale dynamics is best understood for excitatory neurons, for simplicity and clarity, we demonstrate the scaling up procedure for excitatory neural fields. The extension of this procedure to include inhibitory neurons is straightforward and will be presented elsewhere.

2.1. **Microscopic dynamics.** We begin by postulating that cell-scale activity for excitatory neurons is governed by the Hodgkin-Huxley equations [Hodgkin and Huxley, 1952, Gerstner et al., 2014, and many others]

$$\frac{du}{dt} = j_L + j_{Nt} + j_{Na} + j_K, \tag{1a}$$

$$j_L = g_P \left( E_{PL} - u \right); \; j_{Na} = g_{Na} m^3 h \left( E_{Na} - u \right); \; j_K = g_K n^4 \left( E_K - u \right), \tag{1b}$$

$$\frac{d\tilde{\xi}(u,t)}{dt} = \frac{\tilde{\xi}_0(u) - \tilde{\xi}(u,t)}{\tau_{\tilde{\xi}}(u)}, \; \text{where} \; \tilde{\xi} = m, n, h. \tag{1c}$$

In equations (1), $u$ is the membrane potential; $j$ denotes ionic currents due to sodium-potassium pump and passive (leaky) transport (L), to neurotransmitters (Nt), and to Na$^+$ and K$^+$ ion channels. The currents are normalized by the capacity of the cell; $E_{PL}$, $E_{Na}$, $E_K$ are the corresponding reversal potentials; $g_{PL}$, $g_{Na}$ and $g_K$ are constants. The gating variables $m$, $n$, and $h$ depend on time and the membrane potential (equations 1c). Their evolution is described as quasi-exponential relaxation to equilibrium values $\xi_0$ with relaxation times $\tau_\xi$.

The Hodgkin-Huxley equations (1) are not complete representation of the cell-scale dynamics of any specific neuron type (for example, calcium ionic currents are not considered). However, they provide a complex-enough cell-scale model for the illustrating the fundamental problem of scaling up to mesoscale and generate a mesoscopic model complex enough to simulate some relevant collective activity dynamics.

## 2.2. **Scaling up considerations.**

For simplicity, we assume that the dynamical equations (1) are the same for all neurons of the same type in a population, and that neurons are differentiated only by the geometry of their connections and their states.

Based on the aspect of the membrane potential evolution shown in figure (1), we assume that the duration of an action potential event is infinitesimal with respect to the mesoscale characteristic time. This suggests that dynamical elements that have characteristic time of an action potential are not be observable at mesoscale. Thus, the goal of the following simplifications is to formulate approximate equations for neuron dynamics (1) *outside action-potential events*. Some of the ideas below follow classical examples of "reduced", low-dimensional equations such as the FitzHugh-Nagumo (FitzHugh, 1955, 1960, 1961, and Nagumo et al., 1962) and Morris and Lecar [1981] models. In the latter, notably, all parameters are experimentally measurable (other approaches for reducing the dimensionality of the dynamical equations 1 were proposed by Ermentrout and Kopell, 1986, Ermentrout, 1996, Hindmarsh and Cornelius, 2005).

We make the following simplifications for the subthreshold regime outside an action potential:

*i) Potassium currents $j_K$ may be neglected.* Outside an action potential, in the subthreshold regime, $j_K \ll j_{Na}$ and $j_K \ll j_P$, and can be neglected. Although the magnitudes of gating variables $n$ and $m$ are comparable, $n$ appears at 4th power in 1b, and $E_K$ is close to the resting potential (see figure 2, Panel b and c).

*ii) Sodium currents $j_{Na}$ are separable.* Figure 2, panels d and e suggests that $\tau_m \ll \tau_h$, and, in the subthreshold regime, $\tau_h$ may be approximated as constant $\tau_h \approx \tau_h(u_{rest})$, independent of $u$. Because $m$ relaxes quickly to equilibrium values, its explicit dependency on time may be ignored, $m = m(u,t) \approx m(u)$. With $\tau_h \approx$ constant, the gating variable $h$ is only a function of time, $h = h(t)$. Therefore, outside an action potential, the equation for the Na$^+$ current may be simplified to

$$j_{Na}(u,t) = h(t)f(u) \text{, where } f(u) = g_{Na}m^3(u - E_{Na}),\tag{2}$$

where $f$ is monotonically increasing with $u$.

*iii) Neurotransmitter-controlled currents depend on the presynaptic neurotransmitter flux $\phi_{Nt}$,*

$$j_{Nt} = j_{Nt}(\phi_{Nt}),\tag{3}$$

where $\phi_{Nt}$ is the amount of presynaptic neurotransmitters received in the unit of time.

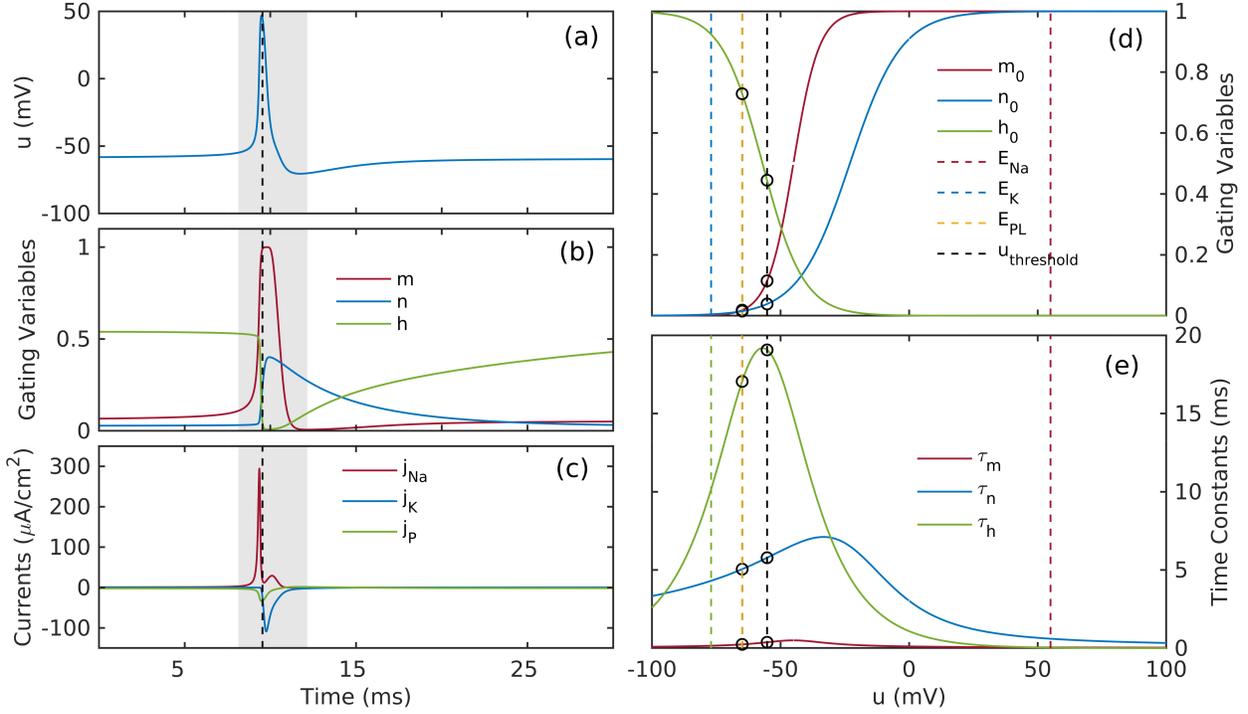

FIGURE 2. Numerical simulations of an action potential using the Hodgkin-Huxley equations 1, with $g_{Na} = 40$ mS/cm$^2$, $g_K = 60$ mS/cm$^2$ and $g_{PL} = 0.3$ mS/cm$^2$. a) Membrane potential $u$. b) Gating variables $m, h, n$. c) Ion currents. The the action potential event is marked by vertical dotted lines. d) Voltage-dependent equilibrium values of gating variables $m, h$ and $n$ (reversal potentials, see equations 1, are marked by vertical lines, ). Circles mark resting ($\sim -65$ mV) and threshold ($\sim -55$ mV) states of the membrane potentials in absence of refractoriness. e) Voltage-dependent time constants of the gating variables. Subthreshold narrow band (from $\sim -65$ to $\sim -55$) is essential in the macroscopic limit. In the macroscopic limit, the duration of the action potential becomes infinitesimal, and the details of its shape are ignorable.

*iv) Threshold criterion is defined based on ionic currents.* Although most integrate-and-fire models assume that action potentials are triggered when the membrane potential exceeds a given threshold (typically given as $\approx -55$ mV), the threshold for potential has a wide PDF and therefore is not well defined (e.g., figure 1; see also not a unique valu Koch et al., 1995, Rinzel and Ermentrout, 1998, Gerstner et al., 2014). This variability is in fact codified in the dynamical equations (equation 1b). Because in the refractory state $h$ is small (figure 2 b-c), the sodium channel is relatively insensitive to the voltage $u$; therefore, the susceptibility of the neuron to fire immediately after an action potential is diminished, i.e., the threshold potential is higher. This effect decreases with time. A threshold criterion based on ionic currents compensates for this effect. Indeed, setting a fixed value $j_c = hf(u)$ (equation 2) for the threshold critical current implies that a higher potential is required for the initiation of an action potential when $h$ is low (see also figure (3)). In this representation, the gating variable $h$ is a measure of cell "excitability". This is consistent with the observation that the current threshold is a better representation for slowly varying input (e.g., Koch et al., 1995, also figure 1). Switching to the current threshold is equivalent to changing the state function from $u$ to $j_{Na}$,

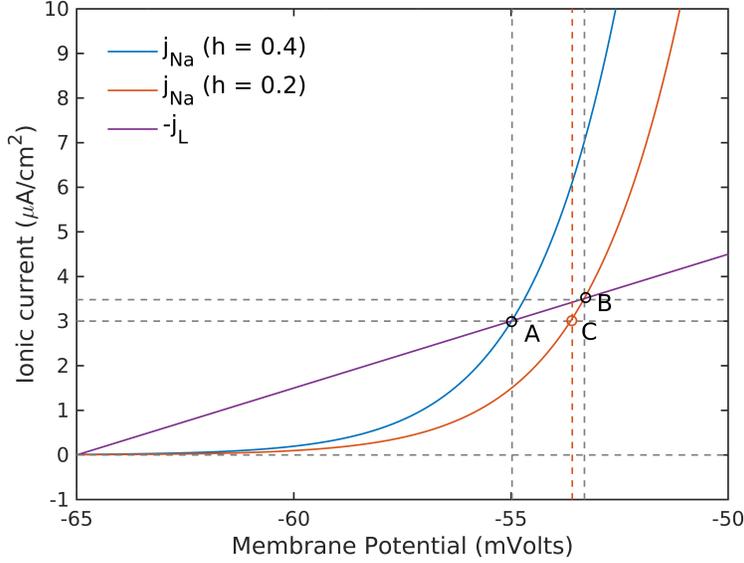

FIGURE 3. Dependency of ionic currents on membrane voltage $u$, for two values of $h$, the lower value representing a refractory state. The threshold criterion may be defined by the unstable fixed points $A$ and $B$, where the sodium current balances the total current due to the combined currents through the pump and leaky channels (purple). The potassium current in in the subthreshold regime is negligible. For different values of $h$, this criterion returns different critical values for both membrane potential, $u^A \neq u^B$, and sodium current $j_{Na}^A \neq j_{Na}^B$. The fixed critical current threshold used in this study, point $C$ (red circle), partially compensates for the difference $u^A - u^B$ due refractoriness. See text for more details.

i.e., inverting equation (2)

$$u = u(j_{Na}, t) = f^{-1}\left(\frac{j_{Na}}{h}\right) \tag{4}$$

(equation (2) is invertible because $f(u)$ is monotonic in the subthreshold regime).

Differentiating equation (2) to time, using equations (1a) and (1c), and neglecting $j_K$ obtains

$$\frac{d}{dt}j_{Na} = hf'(u)j_{Nt} + hf'(u)(j_L + j_{Na}) + \frac{1}{\tau_h}f(u)(h_0 - h), \tag{5}$$

where all dependencies on $u$ are in fact dependencies on $j_{Na}$ through the composite function $u(j_{Na}, t)$, and $j_P = j_P(u(j_{Na}, t))$ via equation (1b). Denoting neurotransmitter-induced ionic channels to the rate of change of $j_{Na}$ by

$$\phi_{Na} = h_0 f'(u)j_{Nt}(\phi_{Nt}), \tag{6}$$

equation (5) may be rewritten in the simpler form

$$\frac{d}{dt}j_{Na} = \frac{h}{h_0}\phi_{Na}(j_{Na}, \phi_{Nt}) - \frac{j_{Na}}{\tau_j(j_{Na}, h)}, \text{ with } \tau_j(j_{Na}, h) = \left[hf'(u)\left(\frac{j_L}{j_{Na}} + 1\right) + \frac{h_0 - h}{h\tau_h}\right]^{-1}. \tag{7}$$

In the first term on the right-hand represents the external forcing, and the second term represents the natural decay of $j_{Na}$, with a state-dependent relaxation time $\tau_j(j_{Na}, h)$.

Replacing equation (1a) with equation (5) obtains the pair of equations

$$\frac{d}{dt} j_{\text{Na}} = \frac{h}{h_0} \phi_{\text{Na}} \left( j_{\text{Na}}, \phi_{Nt} \right) - \frac{j_{\text{Na}}}{\tau_j \left( j_{\text{Na}}, h \right)}; \; \frac{d}{dt} h \left( j_{\text{Na}}, t \right) = \frac{h_0 - h \left( j_{\text{Na}}, t \right)}{\tau_h \left( j_{\text{Na}}, h \right)}. \tag{8}$$

Although equations 8 describe the subthreshold regime they are dynamically incomplete because eliminating the detailed dynamics of the action potential also eliminates the effect of resetting to resting state immediately after action potential. This may be corrected by introducing a Heaviside function (in differential form, Dirac delta function) that returns $j_{Na}$ and $h$ to 0 when $j_{\text{Na}} = j_c$, where $j_c$ is the critical threshold current for the initiation of an action potential (e.g., Roxin et al., 2011). The complete differential equations for subthreshold dynamics become

$$\frac{d}{dt} j_{\text{Na}} = \frac{h}{h_0} \phi_{\text{Na}} \left( j_{\text{Na}}, \phi_{Nt} \right) - \frac{j_{\text{Na}}}{\tau_j \left( j_{\text{Na}}, h \right)} - j_{\text{Na}} \delta \left( j_{\text{Na}} - j_c \right); \tag{9}$$

$$\frac{d}{dt} h = \frac{h_0 - h \left( j_{\text{Na}}, t \right)}{\tau_h \left( j_{\text{Na}}, h \right)} - h \delta \left( j_{\text{Na}} - j_c \right). \tag{10}$$

Figure (4) compares the behavior of the reduced model (9) comparing to the Hodgkin-Huxley model (1) under random current injections. As expected, the reduced model does not capture the membrane activity in within the duration of an action potential, but represents the evolution of in subthreshold regime very close to the Hodgkin-Huxley model; note that the reduced model ignores the contribution of the of potassium channels.

*Cell activity as energy balance: the "powder keg" model.* Just like the Hodgkin-Huxley model (1), the reduced model (9-10) has a clear interpretation in terms of energy balance, often expressed in electric diagrams, [e.g., Gerstner et al., 2014]. As in any electric circuit, $u$ is potential energy stored by capacitors, created and maintained by the work done by ion pumps to separate charges, while ion currents represent kinetic energy carried by moving charges. Neuronal dynamics is fundamentally a conversion between potential and kinetic energy, controlled by highly variable conductivity rates. The energy stored is maintained through a sisyphean balance between two opposite processes: ion pumps do work to create/maintain the stored potential against continuous losses through conversion to kinetic energy. Due to variable conductivity rates (gating variables), if the kinetic energy exceeds a threshold, the balance is lost and all energy stored is released explosively to kinetic energy during the action potential[4]. Note that action potentials (and related firing rates) in fact characterize the energy input into neural field, and not the state of the field.

Membrane potential fluctuations are externally driven by changes in presynaptic neurotransmitter concentration which modulate neurotransmitter-controlled ionic currents. A neuron is an energy storage unit, equipped with a sensor for "ambient" neurotransmitter concentration, and rigged to release explosively the stored energy when ionic currents reach a threshold value. The net effect of the energy release is to increase presynaptic neurotransmitter concentration at connected neurons.

This representation is strikingly similar to a powder keg rigged to detonate when the ambient temperature reaches a given threshold. Black powder stores potential energy; releasing the potential energy stored in the powder converts it to kinetic energy, a fraction of which in the form of heat, i.e., internal kinetic energy, that raises the ambient temperature at nearby kegs (temperature and chemical potential of neurotransmitters are both internal energy densities, e.g., normalized by number of carrying particles). The details of energy conversion events (action potentials) have no dynamic relevance[5]; the fundamental role of action potentials is to convert membrane potential energy into kinetic energy (ionic currents) at connected neurons, by modulating their presynaptic neurotransmitter chemical potential.

---

[4]Some of the energy is lost in various forms, e.g., electromagnetic radiation, or simply work on milieu ions.

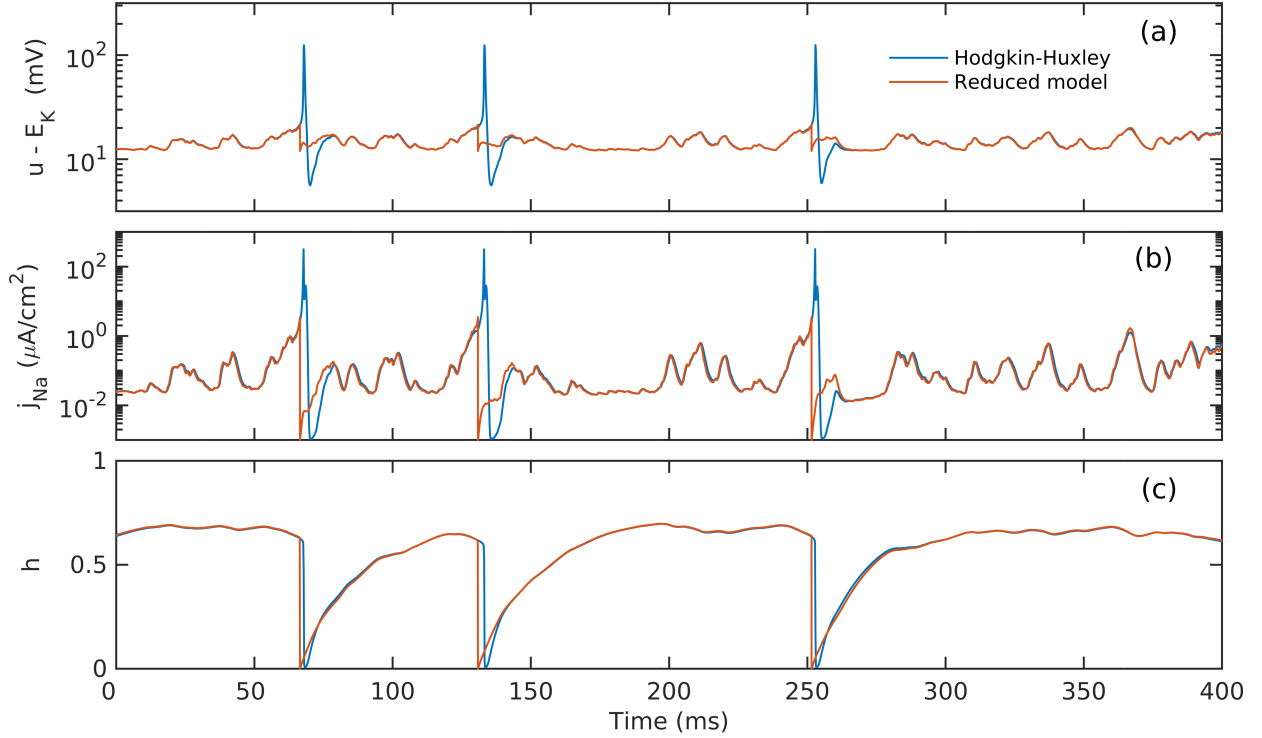

FIGURE 4. Evolution of membrane potential as described by the Hodgkin-Huxley model (equations 1; blue) and the subthreshold reduced model (equations 9-10; red) under identical forcing by random current. a) Membrane potential $u$; b) sodium current $j_{Na}$; c) gating variable $h$.

2.3. **Mesoscopic dynamics, a heuristic model.** The mesoscopic governing equations are obtained by averaging the dynamical equations (9) over a population of neurons. Below, we assume for simplicity that a) although the geometry of connections may be different for each cell, the average connection configuration is homogeneous and isotropic across the neural field; b) the lag between an action potential and neurotransmitter release is negligible at mesoscale; however, c) the characteristic neural connection distance $\Lambda$ is a mesoscale length, small, but finite (connections are local). Let $x = \{x_i\}$ and $t$ denote the mesoscopic spatial position vector and time, with $x_i$, $i = 1, 2, ...$ denoting spatial coordinates. Using the notation for scale order introduced in section (1), we write

$$\text{cell scale} \ll dx < \Lambda < \text{neural field diameter,} \tag{11}$$

---

[5] This perspective is not dissimilar to molecule collisions in a gas: each collision event is different and has complicated details, but the net result is a redistribution energy across the molecule population. Note that action potentials inject *new energy* into the neural field (this is distinctive of excitable systems).

where $dx$ is the mesoscopic element of length. A mesoscopic quantity $Q$ is defined here by applying the averaging operator $Q = \langle q \rangle$ to microscopic variable $q$

$$\langle q \rangle = \lim_{V \to 0} \frac{1}{\rho V} \sum_{\text{cells in } V} q; \text{ where } \rho = \lim_{V \to 0} \frac{n_{cells}}{V}, \tag{12}$$

where $n_{cells}$ is the number of cells in the volume $V$, and $\rho$ is the local cell density. Applying the average operator (12)) to equations 9, assuming that $\tau_j(j, h) \approx \tau_j$ constant, and $h$ and $\phi_{\text{Na}}$ are uncorrelated obtains

$$\frac{d}{dt} J = H\Phi - \frac{1}{\tau_j} J - N j_c; \tag{13a}$$

$$\frac{d}{dt} H = \frac{1}{\tau_h} (1 - H) - N \left( \frac{h_N}{h_0} \right), \tag{13b}$$

where $J$, $H$, $\Phi$ and $N$ are mesoscopic quantities corresponding to sodium current, "excitability", "flux of excitation", and firing rate (number of firing events per unit time and volume, normalized by cell density)

$$J = \langle j_{\text{Na}} \rangle, \ H = \left\langle \frac{h}{h_0} \right\rangle, \ \Phi = \langle \phi_{Nt} \rangle, \ N = \lim_{\Delta t, V \to 0} \frac{\sum_{\text{cells in} V} \int_t^{t+\Delta t} \delta \left( j_{\text{Na}} - j_c \right) dt}{\rho V \Delta t}. \tag{14}$$

The source of presynaptic neurotransmitters at a fixed point $x$ of the neural field may be written as

$$\Phi = \left\langle \phi_{\text{Na}} \left( j_{\text{Na}}, Q(x, t) + \int_{-\infty}^{\infty} N(\zeta, t) F(\zeta, x) d^3 \zeta \right) \right\rangle, \tag{15}$$

where $\int_{-\infty}^{\infty} N(\zeta, t) F(\zeta, x) d^3 \zeta$ and $Q(x, t)$ represent neurotransmitter influx at $x$ due to internal and external neural activity, respectively. The neurotransmitter flux of excitation is characterized by both its origin and destination. The influx from activity within the field is expressed by an integral of the firing rate over the entire field, weighted by the distribution $F(\zeta, x)$ of connections from element of volume $d^3 \zeta$ at $\zeta$ to element of volume $d^3 x$ at $x$. The distribution $F(\zeta, x)$ is normalized, in the sense that the zeroth moment, "number" of connections to $x$ from the entire neural field, is $M_0 = \int_{-\infty}^{\infty} d^3 \zeta \ F(\zeta, x) = 1$. The excitation flux from sources external to the neural field, however, is a forcing term; the details of its origin are irrelevant, therefore in contains only destination information.

Note that equation (15) is symbolic, in the sense that the exact form of the dependency of $\phi_{\text{Na}}$ on excitation involves complex biological connectivity factors such as the geometry of dendrites, dynamics of neuroreceptors, and others. To convert it to a tractable form, we make below a few a simplifying assumptions.

If neural connectivity is assumed homogeneous, isotropic, and local, then the form of the dependency $\Phi(N)$ is invariant to translations and may be reduced to local values. The distribution of connection is $F(\zeta, x)$ depends only on the distance between points $x$ and $\zeta$, (symmetric in $x$ and $\zeta$), i.e., $F(\zeta, x) = F(x, \zeta) = F(\|\delta\zeta\|)$, where $\delta\zeta = \zeta - x$, with components $\delta\zeta_n = \zeta_n - x_n, n = 1, 2, 3$. Expanding the integral $\int_{-\infty}^{\infty} N(\zeta, t) F(\zeta, x) d^3 \zeta$ in a Taylor series at $x$, keeping only the leading-order spatial terms (local connections means that $F$ decays fast enough with $\|\delta\zeta\|$, and accounting for the symmetries of various moments of $F(\|\delta x\|)$ obtains

$$\int_{-\infty}^{\infty} N(\zeta, t) F(x, \zeta) d^3 \zeta = M_0 N + M_2 \nabla^2 N + O \left( \|\delta\zeta\|^4 \right), \tag{16}$$

where $M_2 = \int_{-\infty}^{\infty} (\delta x_n)^2 F(\|\delta x\|) d^3 \zeta$ is the second moment of $F$. The excitation flux incoming at $x$ from internal action potentials is then $\Phi \simeq \Phi (M_0 N + M_2 \nabla^2 N + Q)$. Finally, we further simply this relationship to a linear one

$$\Phi \simeq \epsilon (M_0 N + M_2 \nabla^2 N + Q), \tag{17}$$

where "connection strength" parameter $\epsilon$ is the average rate of change of the sodium current $J$ caused by a unit firing rate, in the absence of refractoriness (first term in equation (13a)). Equations (13a) take now the form

$$\frac{d}{dt} J = \epsilon H (M_0 N + M_2 \nabla^2 N + Q) - \frac{1}{\tau_j} J - N j_c, \tag{18}$$

$$\frac{d}{dt} H = \frac{1}{\tau_h} (1 - H) - N \left( \frac{h_N}{h_0} \right). \tag{19}$$

Equation (18) represents the kinetic energy rate of change as balance between production by action potentials (first term, with $\nabla^2 N$ describing the spatial the distribution of energy sources) and natural dissipation mainly caused by the leaky channels (second term). The last term describes the decay associated with the resetting effect of action potentials. Equation (19), describes the rate of change of the excitability of the system as a balance between natural recovery, associate with the natural tendency of gating variable $h$ to trend towards the resting state, and the loss caused by firing events.

## 2.4. The activation function.

The activation function links the local firing rate to the state space of the local cell ensemble in the element volume. At cell scale, the membrane potential evolution itself is a stochastic process that emerges as an integration of random ion currents produced by a large population of diverse ion channels. This suggests that the evolution of $j_{Na}$ at cell scale may be represented as an Itô integral of a standard Wiener process [e.g., Kleinert, 2009, Karatzas and Shreve, 2012, Cohen and Elliott, 2015]

$$dj_{Na} = (f_E - c_L j_{Na}) dt + \sigma dW_t, \tag{20}$$

where $\sigma dW_t$ represents the strength of the intrinsic membrane fluctuation with a diffusion coefficient $D = \sigma^2/2$, and the drift coefficients $f_E$ and $c_L j_{Na}$ represent the opposite effects of forcing by excitation (or inhibition) caused by neurotransmitter, and decay due to leaky ion currents. The incoming energy flux determines the averaged trend of membrane depolarization while endogenous fluctuation (the random walk) determines oscillatory deviation from the trend. For simplicity, we assume here that the forcing drift is independent of $j_{Na}$, the decay drift is independent of excitation, and $c_L$ is constant. The probability density function $P(j_{Na}, t)$ of the sodium current over the neuron population has the properties that $P(j_{Na} = j_c, t) = 0$ and $\int_0^{j_{Na}^c} P(j_{Na}, t) = 1$, and satisfies the Fokker-Planck equation [e.g., Risken, 1989]

$$\frac{\partial P(j_{Na}, t)}{\partial t} = D \frac{\partial^2 P(j_{Na}, t)}{\partial j_{Na}^2} + \frac{\partial [(f_E - c_L j_{Na}) P(j_{Na}, t)]}{\partial j_{Na}}. \tag{21}$$

In this formulation, the mesoscale firing rate $N$ (number of spikes/time/number of cells in unit volume) is the probability flux across the critical threshold value $j_c$, given by equation (20),

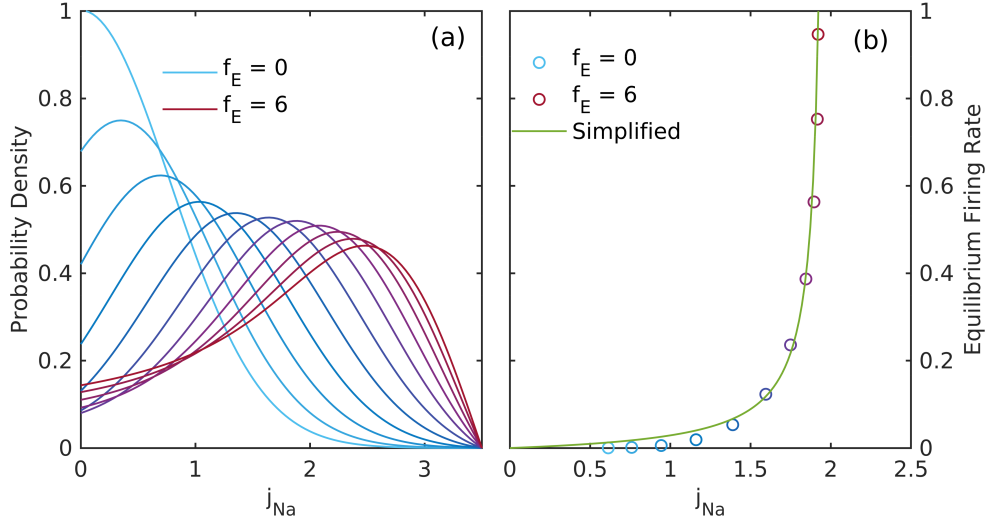

FIGURE 5. a) Equilibrium solutions of equation (24) for various values of forcing due to excitation $0 < f_E < 6$ ($f_E$ values increase from blue to red). b) Firing rate $N$ as a function of kinetic energy density (temperature) $J$. Values derived from the equilibrium solutions of equation (24) are marked by circles. The green line is the activation function, equation (25).

$$N(t) = -D \left. \frac{\partial P(j_{Na}, t)}{\partial j_{Na}} \right|_{j_c}.$$ (22)

If the forcing excitation drift $f_E$ is considered constant, the stationary solution satisfies

$$D \frac{d^2 P(j_{Na})}{dj_{Na}^2} + \frac{d\left[(f_E - c_L j_{Na}) P(j_{Na})\right]}{dj_{Na}} = 0,$$ (23)

and has the form

$$P(j_{Na}) = A \sqrt{\frac{\pi D}{2c_L}} e^{\frac{-(f_E - c_L j_{Na})}{2Dc_L}} \, \text{erfi}\left(\frac{c_L j_{Na} - f_E}{\sqrt{2Dc_L}}\right) + B e^{\frac{(f_E j_{Na} - \frac{1}{2} c_L j_{Na}^2)}{D}}$$ (24)

where $A$ and $B$ arbitrary constants. The solution (24) for various values of the forcing input $f_E$ values is shown in figure (5a). The theoretical *static* activation function, i.e., the relation (22) at equilibrium states is shown figure (5b). For simplicity, we represent here the relationship between the mesoscopic firing rate and mesoscopic kinetic energy $J$ by the analytic function

$$N(J) = A \left(\frac{1}{J_c - J} - \frac{1}{J_c}\right),$$ (25)

which approximates well the static activation function (figure 5 b). In equation 25, $J_c$ is a constant that controls the growth of the firing rate in the vicinity of the threshold value $j_c$; they describe the similar threshold process, the former at cell scale and the latter at mesoscale; therefore, they are related, but not equal. The constant $A$ controls the response of the firing rate to to the kinetic energy density of the field $J$, and will be referred to as mean-neuron "susceptibility" (note that $A$ plays the role of diffusion coefficient in the Fokker-Planck formulation 21).

2.5. **Summary of the collective activity model.** The complete model proposed here for mesoscopic activity in an excitatory neural field is given by equations

$$\frac{\partial}{\partial t} J = \epsilon H \left( M_0 N + M_2 \nabla^2 N + Q \right) - \frac{1}{\tau_j} J - N j_c, \tag{26a}$$

$$\frac{\partial}{\partial t} H = \frac{1}{\tau_h} \left( 1 - H \right) - N \frac{h_N}{h_0}, \tag{26b}$$

with a simplest form of activation function

$$N(J) = A \left( \frac{1}{J_c - J} - \frac{1}{J_c} \right). \tag{26c}$$

In this formulation, the state of the neural field is completely determined by two state variables, $J$ and $H$.

The neural field configuration is characterized by four parameters: $\tau_j$ – the time constant of kinetic energy; the refractory time constant $\tau_h$; the connection strength $\epsilon$; and the mean-neuron susceptibility $A$. While these parameters may vary in time and space, we will assume below that their characteristic scales are much larger that mesoscale. For simulation purposes, they are considered here constant.

The state of the neural field is described by two parameters: the kinetic energy density $J$, and the excitability $H$. The variables $N$ (firing rate) and $Q$ (external forcing) describe the energy injected into the system by action potentials, and by external sources, respectively, therefore they are process variables. Both state and process variables are intensive (densities, i.e., normalized by the number of cells in the unit volume).

The analogy with powder-keg dynamics also extends to the mesoscale description (26). Consider in this case field of powder-kegs: the kinetic energy density $J$ is equivalent to the local temperature of the powder-keg field; the local temperature of the field increases due to explosions ($N$) and heat received from outside ($Q$), and decay naturally due to radiation leakage ($\tau_j$); the excitability $H$ is a measure of powder-keg availability – imagine that every exploded keg takes some time to replace; etc. *Because the relationship between $J$ and the concept of temperature is particularly intuitive, below, we will refer to $J$ as the "field temperature" of the neural field.*

Equations (26) are complete in the sense that they provide enough information to determine the future state of the system based on the current state and the energy exchange between the system and the environment. Although the model (26) is nonlinear (e.g., first term in equation (26a), second term in equation activation function, (26b), and the activation function itself, equation (26c)), because the linear properties of the equations can play an important role in the behavior of a nonlinear system, in the remainder of this study we focus on investigating the properties of the linearized .

## 3. Equilibrium and stability of the linearized system

Below, the neural field is assumed to be under a steady, spatially uniform input, i..e., $\frac{\partial Q}{\partial t} = 0$ and $\nabla Q = 0$. We normalize $J$ and $\tau_j$ by $J_c$ and $\tau_h$ respectively for analytical solutions, i.e.

$J_c = \tau_h = 1$ and we assume that the cells fire are mostly the cells in their resting states for simplicity such that $h_N = h_0$.

To describe perturbations around stable equilibrium states that that may vary in space, let $\delta \ll 1$ be a small parameter that measures the magnitude of the departure from equilibrium states, and expand the state variables in the asymptotic series

$$J = J_0 + \delta J_1 + O\left(\delta^2\right); \quad H = H_0 + \delta H_1 + O\left(\delta^2\right), \tag{27}$$

where the zero-subscripts denote the equilibrium states. Process variables $N$ and $\Phi$ are also expanded in asymptotic series

$$N = N_0 + \delta N_1 + O\left(\delta^2\right); \; N_0 = N(J_0); \; N_1 = \frac{\partial N}{\partial J} J_1, \tag{28}$$

$$\Phi = \Phi_0 + \delta \Phi_1 + O\left(\delta^2\right); \; \Phi_0 = \Phi(N_0); \; \Phi_1 = \frac{d\Phi}{dN} N_1, \tag{29}$$

where $\Phi$ is treated as a functional of $N$, and $\frac{d\Phi}{dN}$ is the variational derivative.

Equilibrium states are defined here by the condition that the kinetic energy of the system is stationary and constant in space, $\frac{\partial}{\partial t}(J, H) = (0,0)$ and $\nabla(J, H) = (0,0)$, therefore, the energy flux and firing rate at equilibrium are homogeneous, e.g., $\nabla^2(N, \Phi) = (0,0)$. Substituting expansions (27- 29) into the governing equations 26 and separating the powers of $\delta$ obtains the standard hierarchy of systems for each power of $\delta$.

3.1. **Equilibrium states.** At $O\left(\delta^0\right)$, the equation for the equilibrium state is

$$\mathcal{F}_{\text{in}} = \mathcal{F}_{\text{out}} \tag{30}$$

where

$$\mathcal{F}_{\text{in}} = \epsilon\left(Q + N_0\right)\left(1 - \tau_h N_0\right), \; \mathcal{F}_{\text{out}} = N_0 j_c + \frac{1}{\tau_j} J_0. \tag{31}$$

with $\mathcal{F}_{\text{in}}$ and $\mathcal{F}_{\text{out}}$ the kinetic energy gains and losses, respectively. Equation 30 describes equilibrium states as a balance of kinetic energy gains and losses achieved for some firing rate $N_0$. Substituting the relation (25) into equation (30) obtains the cubic algebraic equation

$$p_3 N_0^3 + p_2 N_0^2 + p_1 N_0 + p_0 = 0, \tag{32}$$

with the coefficients

$$p_0 = A\left(\epsilon Q - \frac{1}{\tau_j} j_c + \frac{1}{\tau_j}\right), \; p_1 = \left(Q\left(1 - A\right)\tau_h + A\right)\epsilon - j_c A - \frac{1}{\tau_j} j_c$$
$$p_2 = \epsilon\left(1 - Q - A\right)\tau_h - j_c, \; p_3 = -\epsilon \tau_h. \tag{33}$$

Firing rates corresponding to equilibrium states are real roots of equation 32. Equation 32 may have one or three real roots, depending on the configuration of the field and external forcing. Figure 6a-c shows the $(\epsilon, A)$ distribution of $J$, $H$, and $N_0$ for one-equilibrium states under input $Q = 0.1$. Each point in the $(\epsilon, A)$ plane represents a different neural field, with a different "physical" configuration. Three-equilibrium states occupy the cusp-like white region.

For fields with low values of $\epsilon$ and $A$ (inefficient connectivity, low susceptibility; lower-left corner of figure 6a), the external forcing $Q$ dominates the energy recaptured from the firing rate $N_0$. The equilibrium states are stable, with mesoscopic activity essentially controlled by the external stimulation $Q$.

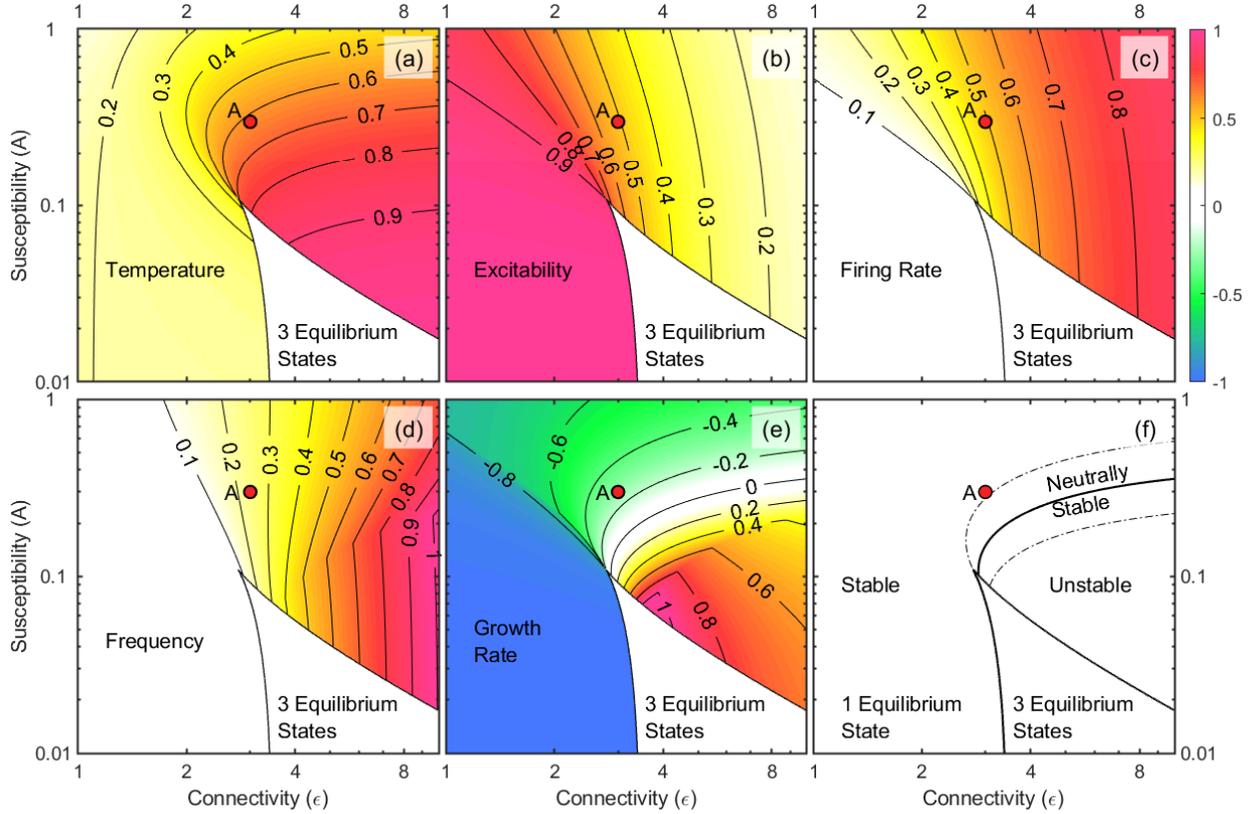

FIGURE 6. Equilibrium states under homogeneous forcing ($Q = 0.1$, $\tau_j = 1.0$) for different field connectivity $\epsilon$ and susceptibility $A$. Each point in the $\epsilon$-$A$ plane represents a different "physical" configuration, i.e., a different neural field. a) Field temperature (or kinetic energy density) $J$. b) Excitability $H$. c) Firing rate $N$. d) Frequency $\omega$. e) Growth rate $\vartheta$. f) A schematic map of the stability of 1-equilibrium states. The neural field configuration to point A (red circle), is used in the analysis presented below.

Alternatively, for fields with high connectivity and susceptibility (upper-right corner of figure 6a), the firing rate $N_0$ contributes a significant amount of energy to the equilibrium state. As $J_0$ and $N_0$ increases, higher order terms in equation (32) play an increasingly important role, the relationship between $(J_0, N_0)$ and external input $Q$ weakens, and the stability of the equilibrium point decreases. Although the equilibrium state becomes locally unstable as $J_0$ and $N_0$ increases, the nonlinearity associated with the modulation of excitability $H$ (equation 26b) maintains the global stability.

Triple equilibrium states exist for fields with low susceptibility $A$ and strong connectivity $\epsilon$ (typically with two stable points separated by an unstable one). Low susceptibility maintains stability at low field temperatures by reducing the effect of endogenous activity on firing rates; strong connectivity $\epsilon$ maintains stability at high field temperatures by sustaining a high firing rate and recapturing a large amount of the energy released. Switching between the two stable states is possible by stimulation or inhibition that can produce large-enough changes in the kinetic energy.

The stability of equilibrium states is described qualitatively in figure 6d-f. Going around the cusp of the three-equilibrium domain, clockwise: frequency grows (panel d), and the growth rate (panel e) increases from negative values (dissipation, stable equilibrium) to positive values (true growth, unstable equilibrium). The oscillatory behavior is confined to the upper right domain in the $(\epsilon, A)$ plane, with self-sustained oscillations approximately confined to the white area in the neighborhood of zero growth (panels e and f). The domains of stability are sketched in panel (f). Figure (6) suggests that mesoscopic oscillations are encountered in highly connected fields (large $\epsilon$), consistent with previous results [e.g., Pinto et al., 2005, Trevelyan et al., 2007]. At large firing rates, e.g., for fields with increased susceptibility $A$, the excitability $H$ variable begins to play a role: as $J$ increase, $H$ decreases significantly, which activates the nonlinear term $H\Phi$ in equations (13a). Because excitability introduces in the dynamics a hysteresis effect [Cowan et al., 2016] with a time scale in order of $\tau_h$, a population reaches its equilibrium field temperature form a deviated state would not coincide with reach of equilibrium excitability. The population will not stop at the equilibrium field temperature but overshoot it. As a result, the field temperature $J$ and excitability $H$ are phase lagged; their interplay provides the support for the oscillatory behavior.

## 3.2. Perturbations of equilibrium.

At $O\left(\delta^1\right)$, the system of equations for the leading order perturbation are

$$\frac{\partial J_1}{\partial t} = \epsilon H_1 \left(Q + N_0\right) + \epsilon H_0 N_1 + \epsilon M_2 H_0 \frac{\partial^2 N_1}{\partial x^2} - N_1 j_c - \frac{1}{\tau_j} J_1, \tag{34}$$

$$\frac{\partial H_1}{\partial t} = -\frac{1}{\tau_h} H_1 - N_1. \tag{35}$$

Equations (34-35) may be used to examine the stability of equilibrium states under homogeneous perturbations, or to study the dynamics of inhomogeneous perturbations (collective action).

### 3.2.1. *Homogeneous perturbations.*

For homogeneous perturbations, $\frac{\partial N_1}{\partial x} = 0$ and equations (34-35) become

$$\frac{dJ_1}{dt} = \epsilon H_1 \left(Q + N_0\right) + \epsilon H_0 N_1 - s_0 J_1 j_c - \frac{1}{\tau_j} J_1, \tag{36a}$$

$$\frac{dH_1}{dt} = -\frac{H_1}{\tau_h} - s_0 J_1, \tag{36b}$$

where $s_0 = \left(\frac{dN}{dJ}\right)_0$. Substituting into equations 36 the standard solution $H = e^{i\sigma t}$, $\sigma \in \mathbb{C}$, where $\omega = \mathfrak{R}\{\sigma\}$ is the frequency of oscillation, and $\vartheta = -\mathfrak{I}\{\sigma\}$ is the growth rate ($-\vartheta$ is the decay rate), obtains

$$\sigma = \frac{1}{2} \left(\frac{1}{\tau_h} - q_1\right) i \pm \frac{1}{2}\sqrt{\Delta}, \ \Delta = 4q_2 - \left(\frac{1}{\tau_h} - q_1\right)^2, \tag{37}$$

where $q_1 = \epsilon H_0 s_0 - j_c s_0 - \frac{1}{\tau_j}, q_2 = s_0 \Phi_0 - \frac{q_1}{\tau_h}$.

Pure growth (decay) behavior occurs if $\Delta \leq 0$ in equation 37, with growth corresponding in general to $\left(\frac{1}{\tau_h} - q_1\right) > 0$, and decay with $\left(\frac{1}{\tau_h} - q_1\right) < 0$. Oscillations require $\omega = \mathfrak{R}\{\sigma\} \neq 0$, i.e., $\Delta > 0$ (i.e. non-zero domain in figure 6b). In the general case where both $\omega \neq 0$ and $\vartheta \neq 0$ growth/decay trends dominate eventually oscillatory behavior. Oscillatory behavior dominates in the neighborhood of the $\vartheta = 0$ (white region near the zero contour line in in 6c). Oscillatory patterns arise due to the interplay between the field temperature $J$ and excitability $H$ of the system. For example, if we ignore excitability variations and set $H = 1$, equation (36a) becomes a first order differential equation with no oscillatory solutions (all coefficients are real), which argues that oscillations of the neural field described by equations (36a-36b) have a fundamental refractory nature, similar to those identified by Curtu and Ermentrout [e.g. 2001], Meijer and Coombes [e.g. 2014]. Such refractory oscillations (if they exist) usually have periods in time scale of $\alpha\tau_h$, where $\alpha$ is a factor. These oscillations possibly corresponds to the gamma to ripple frequency range for non-principal cells (e.g. if $3 \lesssim \tau_h \lesssim 6$ ms, $2 \lesssim \alpha \lesssim 4$ then the oscillatory frequency is roughly in-between 42 Hz to 167Hz).

### 3.3. Inhomogeneous perturbations.
Collective activity was defined in section 1 as mesoscopic spatio-temporal patterns of neural activity. In the linear approximation of equations (34-35) collective activity is represented by inhomogeneous perturbations of equilibrium states. Below, we derive their dispersion relation, which completely characterizes their linear dynamics.

If the perturbations have a non-trivial spatial structure, the spatial gradients have to be taken into account. Equations (34-35) may be simplified to retain the field temperature $J$ as the only independent variable, which results in a partial differential equations in $J_1$. Substituting the elementary solution $J_1 \propto e^{i(kx + \sigma t)}$, where $\omega = \Re\{\sigma\}$ is the angular frequency, $\Re\{k\}$ is the wavenumber, and $\vartheta = \Im\{\sigma\}$ and $\Im\{k\}$ are the temporal and spatial growth, rates reduces the equation to the dispersion relation, and algebraic equation $\mathcal{P}(\sigma, k) = 0$, where $\mathcal{P}$ is a polynomial in $\sigma$ and $k$ (Whitham, 2011; see appendix A), which may be solved to provide the $\sigma(k)$

$$\sigma = \frac{1}{2}\left(s_0 H_0 \epsilon M_2 k^2 + \frac{1}{\tau_h} - q_1\right)i \pm \frac{1}{2}\sqrt{4\left(q_2 + \frac{H_0 \epsilon M_2 s_0}{\tau_h}k^2\right) - \left(s_0 H_0 \epsilon M_2 k^2 + \frac{1}{\tau_h} - q_1\right)^2}$$

(38)

where $q_1 = \epsilon H_0 s_0 - j_c s_0 - \frac{1}{\tau_j}, q_2 = s_0 \Phi_0 - \frac{q_1}{\tau_h}$. For progressive waves with $k = \Re\{k\}$, a graphic representation of the solutions is shown in figure (7), for the equilibrium state A (red circle) in (6).

Typical dispersion curves are shown in figure (7). The decay rate monotonically increases with the wavenumber, due to the term $s_0 H_0 \epsilon M_2 k^2$ in equation (38), which is positive for an excitatory neural field. Therefore, in an excitatory neural field, shorter waves have higher frequencies, propagate slower and decay faster. In general, stronger input $Q$ generates higher-frequency waves. Progressive waves satisfying the dispersion relation (38) emerge as an effect of spatial coupling, and are related to "refractory" homogeneous oscillations, the latter being just the degenerate case of zero wavenumber corresponding to non-zero frequency. Because the lower bound of refractory-wave frequency is the frequency of refractory oscillations, the actual values of refractory wave frequencies is likely above the range of cortical and hippocampal ripple frequencies [Buzsáki, 2015] for non-principal cells. This is consistent with the practice of detecting cortex regions with high activity by the LFP power in frequency bands associated with ripples [Ray and Maunsell, 2011].

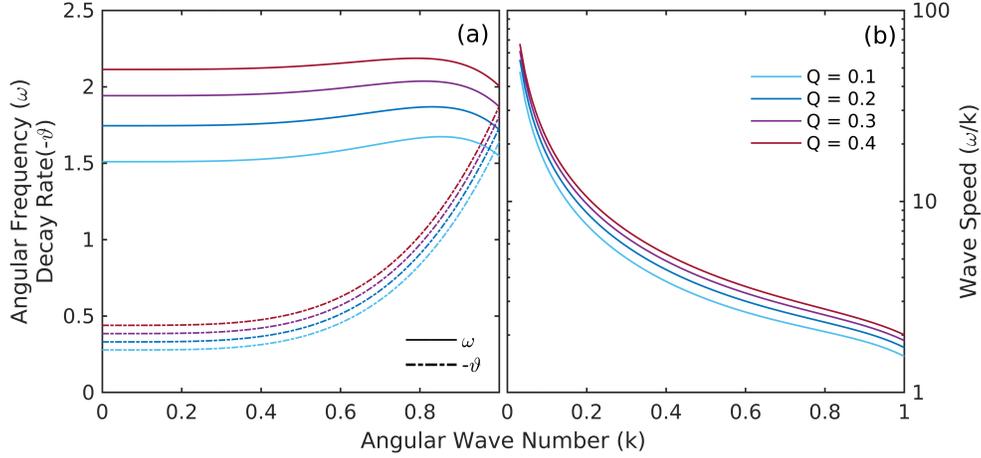

FIGURE 7. The dispersion relation $\sigma(k)$ for $k = \Re\{k\}$ (equation 38) for different values of stimulus forcing $Q$, and field parameters $\epsilon = 3.0$, $A = 0.4$ and $\tau_j = 1.0$ (red dot A in figure 6). a) Angular frequencies and decay rates. b) Non-dimensional phase speed $\omega/k$.

It is interesting to note that, in a neural field of inhibitory cells, because the term $s_0 H_0 \epsilon M_2 k^2$ changes sign ($\epsilon < 0$), the decay rate becomes a monotonically decreasing function of wavenumber and Turing instabilities are possible [Turing, 1990, Métivier and Rauch, 2010], which may indicate a potentially mechanism supporting replay of spatial patterns and memory devices (e.g. Lee et al., 1994, Vanag and Epstein, 2001, 2004, Kaminaga et al., 2006).

## 4. DISCUSSION

This study presents a heuristic formulation of the governing equations for mesoscopic collective activity in the cortex, proceeding directly from generic Hodgkin-Huxley equations for microscopic cell dynamics. We argue that recognizing mesoscopic neural activity as a phenomenon with its own distinct scale intermediate (meso) between cell scale and the global brain scale implies that its governing physical laws governing are distinct. Mesoscale processes are macroscopic with respect to cell-scale activity, and emerge as the "average" behavior of a large population cells. This implies that: 1) the details of dynamics of any single cell are irrelevant; 2) the number of variables that characterize the state of the population is much less than the number of variables necessary to describe the dynamics of every cell of the population.

Based on these assumptions, we demonstrate a simple scaling up procedure for deriving the governing equations for mesoscopic collective activity starting from the dynamical equations describing cell-scale processes. Our approach is highly simplified version of standard statistical mechanics methods for formulating non-equilibrium evolution equations for large populations (hydrodynamics, and the Ising model for magnetism, are examples of fully "worked out" applications ; Kardar, 2007b,a). For simplicity, we postulate that cell-scale dynamics are governed by the Hodgkin and Huxley [1952] equations, and confine our discussion to excitatory neural fields.

The derivation hinges on the observation that, because of their duration, action potential are not observable at mesoscale. While they are essential for cell-scale dynamics, their mesoscopic function is to inject presynaptic neurotransmitters in the neural field thus play a role of kinetic energy redistribution. At mesoscale, the details of an action potential event are irrelevant; what is relevant, is the amount of kinetic energy (e.g., carried by sodium currents) that is created at postsynaptic cells. The macroscopic character of collective activity with respect to cell-scale processes is expressed by equations formulated in continuum form, resulting in a field model. Although heuristic, the scaling up procedure allows for identifying two mesoscopic state variables: the average kinetic energy density $J$ carried by sodium ionic currents, and the excitability of the neural field $H$, which presents the average state of gating variable $h$. Both variables are normalized by number of cells, therefore are, in thermodynamics terminology, intensive variables. Because the definition of $J$ parallels that of the temperature of an ideal gas, we refer to it as "field temperature".

Because the fundamental assumptions about mesoscopic physics made here are the same as in Qin et al. [2020] (e.g., at mesoscale, action potential details are not observable and the exact state of specific individual cell does not matter) the resulting models are similar. However, the equations are not identical, because the more rigorous derivation naturally identifies different state variables (the kinetic energy density $J$ and excitability $H$, in contrast to average membrane potential and a recovery variable). The explicit derivation steps allow for closer inspection, analysis and, corrections, and enhancements of the capabilities of the model.

The resulting neural field model has similarities to WC-class models, but also significant differences: our model is derived directly from cell-scale dynamics (Hodgkin-Huxley equations); the new representation of mesoscopic activity as essentially a subthreshold process; the reassessing of the firing rate as a process variable describing energy transfers (rather than state variable); the natural emergence of the two essential state variables, the field temperature $J$ and excitability $H$; and alternative interpretation of the activation functions.

A preliminary investigation of the linearized collective activity model shows that its properties are consistent with expectations for the dynamics of excitatory neural fields. The system supports oscillations of progressive waves. Stronger stimuli elicit higher frequency responses in both oscillations and waves. Shorter waves typically have higher frequencies, they propagate slower and decay faster, in an excitatory neural field. Although this aspect is not studied here, the model suggest that negative values of connectivity (inhibitory population) support Turing instability and spatial pattern formation (potentially a mechanism for memory Lee et al., 1994, Vanag and Epstein, 2001, 2004, Kaminaga et al., 2006).

This study demonstrates the simplest principles of scaling up from microscopic dynamical equations to obtain the governing equations for macroscopic populations, applied to the simplest formulation of cell dynamics, and to the simplest type (excitatory) of neural field. A more rigorous approach, going from dynamical equations to a statistical formulation, to kinetic Boltzmann equations and finally to the hydrodynamic approximation, is possible and under development, but outside the scope of this study. To be useful for applications, a model of mesoscopic brain activity should also account for complexities such as multiple-type neural fields, including inhibitory neurons, inhomogeneous and anizotropic fields, and others. Research into these aspects is still ongoing.

An analysis of some of the nonlinear properties of the collective activity model, with a focus on cross-scale (frequency) coupling, of relevance to the phenomenon of theta-gamma coupling, is presented in a companion study.

**Compliance with Ethical Standards.** Disclosure of potential conflicts of interest: The authors declare no conflict of interest in the research described in this paper.

Research involving Human Participants and/or Animals: No human or animals were involved in the research described in this paper.

Informed consent: Not applicable.

### APPENDIX A.  DISPERSION RELATION, EQUATION (38) AND EIGENVALUES OF HOMOGENEOUS PERTURBATIONS, EQUATION (37)

To transform the system of equations (34-35) into a single equation in $J$, rewrite equation (35) as

$$J_1 = -\frac{H_1}{s_0 \tau_h} - \frac{1}{s_0}\frac{\partial H_1}{\partial t}$$

and substitute into equation (34). Some algebra obtains the partial differential equation

$$H_0 \epsilon M_2 \frac{\partial^3 J_1}{\partial x^2 \partial t} + \frac{H_0 \epsilon M_2}{\tau_h}\frac{\partial^2 J_1}{\partial x^2} - \frac{1}{s_0}\frac{\partial^2 J_1}{\partial t^2} - \left(\frac{1}{s_0 \tau_h} + j_c + \frac{1}{s_0 \tau_j} - H_0 \epsilon\right)\frac{\partial J_1}{\partial t}$$

$$- \left(\epsilon Q + \epsilon N_0 + \frac{j_c}{\tau_h} + \frac{1}{s_0 \tau_h \tau_j} - \frac{H_0 \epsilon}{\tau_h}\right)J_1 = 0. \tag{39}$$

The dispersion relation is obtained by substituting into equation 39 the elementary solution $J_1 \propto e^{i(kx+\sigma t)}$, where $\omega = \Re\{\sigma\}$ is the angular frequency and $\Re\{k\}$ is the wavenumber, and $\vartheta = -\Im\{\sigma\}$ and $-\Im\{k\}$ are temporal and spatial growth rates. With the derivatives given by the rules $\frac{\partial^n}{\partial t^n} = (i\sigma)^n$ and $\frac{\partial^n}{\partial x^n} = (ik)^n$ one obtains the dispersion relation [Whitham, 2011] as an algebraic equation

$$H_0 \epsilon M_2 \left(ik^2 \sigma\right) + \frac{H_0 \epsilon M_2}{\tau_h}k^2 - \frac{1}{s_0}\sigma^2 + \left(\frac{1}{s_0 \tau_h} + j_c + \frac{1}{s_0 \tau_j} - H_0 \epsilon\right)i\sigma$$

$$+ \left(\epsilon Q + \epsilon N_0 + \frac{j_c}{\tau_h} + \frac{1}{s_0 \tau_h \tau_j} - \frac{H_0 \epsilon}{\tau_h}\right) = 0. \tag{40}$$

Sorting the terms obtains the quadratic equation

$$\sigma^2 - 2is_0 \left(H_0 \epsilon M_2 k^2 \sigma + \frac{1}{s_0 \tau_h} + j_c + \frac{1}{s_0 \tau_j} - H_0 \epsilon\right)\sigma$$

$$-2s_0 \frac{H_0 \epsilon M_2}{\tau_h}k^2 - 2s_0 \left(\epsilon Q + \epsilon N_0 + \frac{j_c}{\tau_h} + \frac{1}{s_0 \tau_h \tau_j} - \frac{H_0 \epsilon}{\tau_h}\right) = 0. \tag{41}$$

Progressive waves are characterized by negligible spatial growth rates; setting $k = \Re\{k\}$, the solutions of equation (41) are

$$\sigma = \frac{1}{2}\left(s_0 H_0 \epsilon M_2 k^2 + \frac{1}{\tau_h} - q_1\right)i \pm \frac{1}{2}\sqrt{4\left(q_2 + \frac{H_0 \epsilon M_2 s_0}{\tau_h}k^2\right) - \left(s_0 H_0 \epsilon M_2 k^2 + \frac{1}{\tau_h} - q_1\right)^2}, \tag{42}$$

where $q_1 = \epsilon H_0 s_0 - j_c s_0 - \frac{1}{\tau_j}, q_2 = s_0 \Phi_0 - \frac{q_1}{\tau_h}$. The last equation is equation (38) in subsection (3.3).

Homogeneous perturbations are a special cases of dispersion relation with $k = 0$. By setting $k = 0$ in 42 obtains 43.

$$\sigma = \frac{1}{2}\left(\frac{1}{\tau_h} - q_1\right)i \pm \frac{1}{2}\sqrt{4q_2 - \left(\frac{1}{\tau_h} - q_1\right)^2},$$ (43)